\documentclass[USenglish,a4paper]{elsarticle2}

\usepackage[USenglish]{babel}
\usepackage[latin1]{inputenc}
\usepackage{amsmath}
\usepackage{amsthm}
\usepackage{amssymb}
\usepackage{multicol}
\usepackage{multirow}
\usepackage{array}
\usepackage{float}
\usepackage{color}
\usepackage{pgf}
\usepackage[ruled,vlined]{algorithm2e}
\usepackage{a4wide}
\usepackage{setspace}
\usepackage{graphicx}
\newcommand{\G}{\mathcal{G}}
\newcommand{\C}{\mathbf{C}}

\newcommand{\rev}[1]{#1}

\begin{document}

\begin{frontmatter}

\title{Generating constrained random graphs using multiple edge switches}
\author[cea,isc,crea]{Lionel Tabourier}\ead{lionel.tabourier@ens-lyon.org}
\author[cams,isc,crea]{Camille Roth}\ead{roth@ehess.fr}
\author[crea,isc,tpv]{Jean-Philippe Cointet}\ead{cointet@polytechnique.edu}
\address[cea]{SPEC, CEA Saclay\\Orme des Merisiers, F-91191 Gif-sur-Yvette, France}
\address[cams]{CAMS, CNRS/EHESS\\54 bd Raspail, F-75006 Paris, France}
\address[isc]{Institute of Complex Systems of Paris-Ile-de-France\\59, rue Lhomond, F-75005 Paris, France}
\address[crea]{CREA, CNRS/Ecole Polytechnique\\ENSTA 32 bd Victor, F-75015 Paris, France}
\address[tpv]{TSV, INRA\\65 Boulevard de Brandebourg\\F-94205 Ivry-sur-Seine Cedex, France}

\begin{abstract}
The generation of random graphs using edge swaps provides a reliable method to draw uniformly random samples of sets of graphs respecting some simple constraints, \hbox{e.g.} degree distributions. However, in general, it is not necessarily possible to access all graphs obeying some given constraints through a classical switching procedure calling on pairs of edges.
We therefore propose to get round this issue by generalizing this classical approach through the use of higher-order edge switches. This method, which we denote by ``$k$-edge switching'', makes it possible to  progressively improve the covered portion of a set of constrained graphs, thereby providing an increasing, asymptotically certain confidence on the statistical representativeness of the obtained sample. 
\end{abstract}


\end{frontmatter}


\section*{Introduction}
The generation of random graphs respecting some constraints has two direct applications: the modeling of realistic network topology when empirical data are missing, and the confirmation of the role of a given set of constraints in the presence of some empirically observed topological and structural features (\hbox{i.e.} some \emph{target observables}, such as in \hbox{e.g.} \cite{newman2004coauthorship}).
There is however currently no general approach to directly create uniformly random graph samples given arbitrary constraints, except for some very specific cases usually related to degree distributions (in this paper, \textit{degree distribution} refers to a specific sequence of degrees, as opposed to a \emph{probability distribution}).

Existing methods for generating random samples of a set of graphs $\G{}_\C{}$ respecting a given set of constraints $\C$ fall indeed into two broad categories:
\begin{itemize}
\item Either by directly building a graph of $\G{}_\C{}$ from scratch, \hbox{i.e.} randomly assigning links to pairs of nodes such that the overall constraint is respected. For instance, the configuration model as presented by \cite{bender1978anl} 
provides random graphs by connecting half-links on nodes such that each resulting graph respects a given prescribed degree distribution.

\item Or by using an original graph which already belongs to $\G{}_\C{}$ and iteratively reshuffling edges of this graph while altogether remaining in $\G{}_\C{}$ in order to asymptotically converge, after a ``sufficient'' number of iterations, to a uniformly random element of $\G_\C$. This approach of switching pairs of edges has been proposed for instance by \cite{rao1996mcm} who aim at obtaining a random graph with a given degree distribution by switching pairs of links in an initial graph which already respects this constraint. \\
The asymptotical convergence is generally empirically appraised with respect to the target observables.
Besides, approaches based on edge swaps implicitly assume that the number of nodes $N$, the number of edges $M$ and the degree sequence are part of $\C$.
In this case, we consider that $\C$ is the union of two subsets: $\C = \C^\emptyset \cup \C^+$, where $\C^\emptyset$ refers to the fundamental constraint forcing graphs to have $M$ links, $N$ nodes, a given degree sequence and to be of a certain type (simple graphs, multigraphs, etc.), while $\C^+$ refers to some additional and arbitrary set of constraints, depending on the context.
\end{itemize}

While the former method assuredly poses a new design challenge for every new kind of constraint --- each set of constraints basically requires a new configuration model --- on the other hand, the latter approach raises the issue of obtaining \emph{uniformly random} elements of $\G{}_\C{}$. Put differently and as we will see below, this reshuffling approach, which initially requires at least one graph from $\G{}_\C{}$, does not guarantee in general that the final graph is \emph{uniformly} chosen from the \emph{whole} set $\G{}_\C{}$.\\

We propose to both (i) appraise the potential issues and drawbacks of random graph creation based on pairwise edge switching (Sec.~\ref{sec:edgeswap}), which is a relatively traditional method in the literature \cite{eggleton1973graphic,colbourn1977gg,tayl:cons,taylor1982scc,rao1996mcm,kannan1997smc,roberts2000sms,milo2003ugr,gkan:mark,stauffer2005ses,PhysRevE.72.056708,viger2005eas,cooper2006srg,feder2006lsm,maha:syst,bansal2008ecr,coolen2009constrained} and, then, (ii) introduce a method for producing random, simulation-based samples of graphs for arbitrary constraints $\C$, using higher-order edge switching processes (Sec.~\ref{sec:kswitch}). 
We eventually present several practical and empirical illustrations in Sec.~\ref{sec:practical}.

\section{Edge swaps as a Markovian reshuffling process}\label{sec:edgeswap}
Mikl\'{o}s \emph{et al.} \cite{miklos2004rpa-annex} showed that it is possible to use a \emph{pairwise edge switching} reshuffling algorithm to generate a uniformly random sample of 
 oriented graphs whose degree distributions are fixed. \cite{PhysRevE.72.056708} later called this method ``switching and holding'' (\emph{S\&H}). 
More precisely, this edge switching method comes to randomly choosing two links in the current graph, checking whether swapping these links leads to a graph respecting the constraint and, if yes, carry out the corresponding swap, otherwise, ``hold'' the current graph and reiterate the procedure. 
 Note that, as such, \emph{S\&H} differs from a simple switching method in that it focuses on the number of swap \emph{trials} rather than the number of swaps.

\begin{figure}
\begin{center}\includegraphics[width=.35\linewidth]{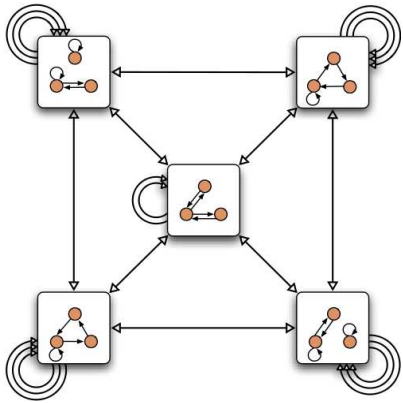}
\end{center}
\caption{\label{fig:metasimple}Simple Markov graph for a constraint on a graph of (i) three nodes with (ii) given in- and out-degree distributions and (iii) without multiple edges but possibly self-loops. Non-valid swaps are represented by self-loops in this Markov graph, which has thus a constant degree.}
\end{figure}

This procedure may be described as a walk in a \emph{Markov graph}. The Markov graph is a directed graph, allowing self-loops and multiple edges such that its set of nodes is exactly $\G_\C$. Arcs of the Markov graph are such that, 
(i) whenever a valid pairwise edge switch transforms $G_i\in\G_\C$ into $G_j\in\G_\C$, we draw an arc from $G_i$ to $G_j$ (and vice versa, mechanically), and (ii) whenever a pairwise edge switch transforms $G_i\in\G_\C$ into a graph which does not belong to $\G_\C$, we draw a self-loop from $G_i$ to $G_i$.
In this context, the reshuffling procedure is a random walk in the Markov graph, that is, a Markov chain \cite{sinclair1993arg} converging to an equilibrium distribution whose probabilities can be obtained from the transition matrix of the Markovian process.
If the Markov graph has constant degrees (\hbox{i.e.} the in-degree and out-degree of all graphs of the Markov graph are all the same), the reshuffling process is uniform. If the Markov graph is connected, all possible graphs are reachable. If it is both connected and has constant degrees, the process leads to uniformly random elements of $\G_\C$. See an illustration on Fig.~\ref{fig:metasimple}. 

Edge switching methods have been used to generate random graph samples in various instances \cite{rao1996mcm,kannan1997smc,stauffer2005ses,cooper2006srg,feder2006lsm,maha:syst,bansal2008ecr} 
and have been studied and improved in various directions \cite{roberts2000sms,milo2003ugr,gkan:mark,PhysRevE.72.056708,viger2005eas}. 
To use such a switching method, one has nonetheless to ensure that all graphs of $\G_\C$ are present in the equilibrium distribution of the random walk with an identical probability, \hbox{i.e.} ensure that:
\begin{enumerate}[(i)]
\item all graphs of $\G_\C$ are \emph{uniformly} drawable, and
\item all graphs of $\G_\C$ are \emph{exhaustively} reachable.
\end{enumerate}

Uniformity is guaranteed by the \emph{S\&H} approach within a given connected portion of the Markov graph. While \cite{miklos2004rpa-annex} show uniformity in the case of degree distribution constraints, the proof they mention in Appendix A of the same reference can easily be extended to any kind of constraint.
A sketch of this proof is given by the following reasoning: 
``holding'' on failed trials is equivalent to connecting a Markov graph node to itself as many times as there are failure possibilities. Thus, the in- and out-degree of all Markov graph nodes will be equal to the number of trials (both failed and successful ones), which is strictly the same for every graph of $\G_\C$, since it only depends on the constant number of links of graphs of $\G_\C$.
Finally, for a random walk in a Markov graph where all nodes have the same in and out-degree, the probability of being on a given node is asymptotically uniform.

Exhaustivity relates to the issue of whether the whole Markov graph is connected, \hbox{i.e.} the existence of a path going from any node to any other node of the Markov graph. In Markov chain terminology, the chain is said to be \emph{irreducible}.
To our knowledge, existing theorems on exhaustivity concern simple constraints $\C$, essentially reduced to little more than the conservation of the original degree sequence: \hbox{i.e.} in the case of trees \cite{colbourn1977gg}, graphs \cite{eggleton1973graphic}, connected graphs \cite{tayl:cons} and bi-connected graphs \cite{taylor1982scc}.

However in the general case of more elaborate constraints (\hbox{e.g.} \cite{maha:syst,bansal2008ecr}), using the \emph{S\&H} method appears to be less legitimate, since no such exhaustivity theorems are known. For instance, Rao \emph{et \hbox{al.}} \cite{rao1996mcm} show that extending $\C$ by requiring the graph to have both directed edges and no self-loop makes it impossible, in some cases, to reach all graphs of $\G_\C$ by pairwise edge swaps. 
In particular, no pairwise edge switch could indeed transform one of the following adjacency matrices into the other one (forbidden self-loops are marked with a star):\\

\begin{footnotesize}
$$ 
\begin{pmatrix}
0^* &1&0 \\
0& 0^* &1 \\
1&0& 0^* 
\end{pmatrix}
\nleftrightarrow
\begin{pmatrix}
0^* &0&1 \\
1& 0^* &0 \\
0&1& 0^*
\end{pmatrix}
$$
\end{footnotesize}

\paragraph{Convergence of the walk}
In addition to these issues, convergence speed remains an open theoretical question \cite{rao1996mcm,guruswami2000rmm}, often coped with using practical heuristics \cite{gkan:mark,viger2005eas}. As said before, the walk usually aims at randomly drawing an element of $\G_\C$ in order to check whether graphs of $\G_\C$ exhibit some properties on the target observables (and, implicitly, in order to check whether $\C$ could constitute a sufficient explanation for these observables). In other words, some measurements are carried out on graphs of $\G_\C$ so that the walk is generally considered to have performed a ``sufficient'' number of steps when those measurements on the target observables apparently plateau.


\section{Higher-order switching process}\label{sec:kswitch}

In this section, for the sake of clarity we focus on directed graphs, although it is effortless to formulate the whole argument for undirected graphs.

\subsection{$k$-edge switching}

In general, the disconnectedness of the Markov graph stems from the impossibility of transforming a graph into another graph by a simple pairwise switching. 
To overcome this issue, we propose an experimental method based on higher-order edge switchings: given $G\in\G_\C$, let us consider $k$ edges $(a_i,b_i)_{i\in\{1,...,k\}}$ corresponding to nodes $(a_1,...,a_k,b_1,...,b_k)$, possibly not all distinct. The $k$-edge switching process, henceforth called ``\emph{$k$-switch}'', comes to randomly choosing one permutation $\sigma$ among the $k!$ possible permutations of $(b_1,...,b_k)$. The resulting graph  is such that edges $(a_i,b_i)_{i\in\{1,...,k\}}$ are replaced with $(a_i,\sigma(b_i))_{i\in\{1,...,k\}}$ (for an example of pseudocode, see Alg.~\ref{fig:algo}). 

It is immediate to see that neighbors of $G$ in the Markov graph corresponding to a classical pairwise edge swap are also neighbors of $G$ in the Markov graph corresponding to a $k$-switch, when considering a permutation that just swaps two $b_i$, $b_{i'}$. Similarly, when $k=2$, we fall back on the \emph{S\&H} approach.

For increasing values of $k$, this procedure creates new links in the Markov graph and new neighbors appear (in the case of Fig.~\ref{fig:metasimple} it is easy to see that the Markov graph is complete for $k=3$). More importantly, some potentially disconnected components of the Markov graph may thus become connected.

\begin{figure}
\begin{center}\includegraphics[width=.65\linewidth]{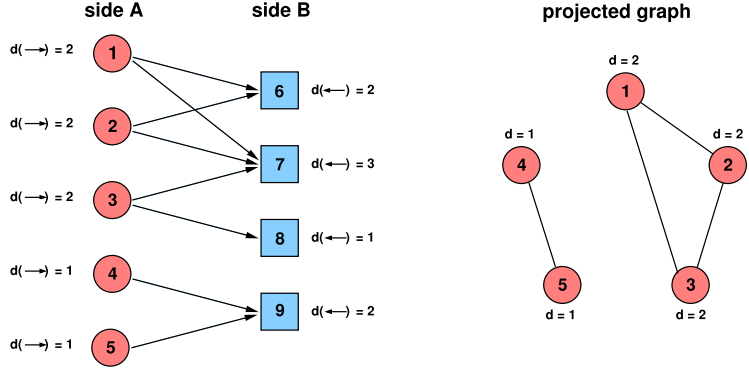}
\end{center}
\caption{\label{fig:ex} \emph{On the left,} one possible realization of a graph drawn from $\G_{C_{0}}$: note that B-sided nodes of the bipartite graph (marked by squares) have out-degree zero and A-sided nodes (marked by circles) have in-degree zero. \emph{On the right,} the corresponding projection of this bipartite graph onto side A.}
\end{figure}

\paragraph{Illustration} 
To illustrate this higher-order switching process, let us consider the case of bipartite (or 2-mode) graphs. Such graphs are useful in the context of real-world networks, for example to study  collaborations in social groups \cite{newman2004coauthorship} or peer-to-peer exchange systems \cite{guillaume2005statistical}. Nodes belong to one of two sides $A$ and $B$, and links connect pairs of nodes from distinct sides only. It is possible to build monopartite (or 1-mode) graphs from the bipartite one by keeping only $A$ (resp. $B$) nodes and linking them if they are connected to the same $B$ (resp. $A$) node in the original bipartite structure, as pictured on Figure~\ref{fig:ex}. These graphs are called \emph{projections} of the original bipartite graph on side $A$ (resp. $B$).\\

Consider a case consisting of a constraint $\C_0=\C^\emptyset_{0}\cup\C^+_{0}$, on bipartite graphs such that:
\begin{quote}\small\em
\begin{enumerate}[(i)]
\item $\C^\emptyset_{0}$: the bipartite graph contains no multiple link, it consists of two sides with fixed degree distributions:
\begin{itemize}
\item ``side A'': 5 nodes, out-degree $\{2,2,2,1,1\}$ (and in-degree 0);
\item ``side B'': 4 nodes, in-degree $\{3,2,2,1\}$ (and out-degree 0).
\end{itemize}
\item $\C^+_{0}$: the degree distribution of the projected graph on side A is fixed: $\{2, 2, 2, 1, 1\}$.
\end{enumerate}
\end{quote}

\begin{figure}
\centering
\includegraphics[width=.8\linewidth]{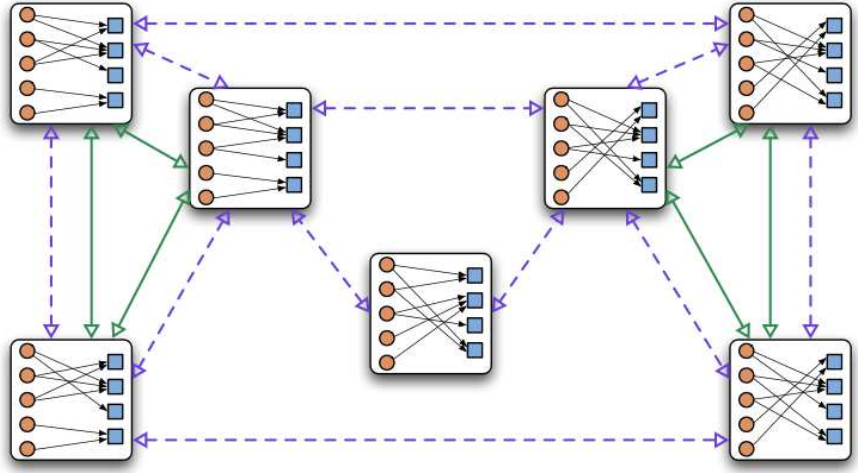}
\caption{\label{fig:metagraph} 
Markov graph of $\G_{\C_0}$ for various $k$-switching procedures: dashed blue arrows correspond to $k=2$, plain green arrows to $k=4$.
For readability purposes, we simplified the representation by discarding self-loops and multiple edges of the Markov graph.}
\end{figure}

Put shortly, this constraint consists in simultaneously imposing degree distributions on a bipartite graph and on one of its monopartite projections. An example of such a graph is represented Fig.~\ref{fig:ex}. Given such a $\C_0$, Markov graphs corresponding to $\G_{\C_0}$ contain $7$ nodes. 
The Markov graph for $k\geq4$ is connected, while it actually consists of $3$ disconnected components for \hbox{$k\in\{2,3\}$} --- see  Fig.~\ref{fig:metagraph}.

We chose this practical case in part because the Markov graph is still small enough to be visualized for each value of $k$. In the remaining examples, it will not be possible anymore,  and no theoretical proof is available; we therefore rely on experimental investigations.

\subsection{Relationship between $k$ and exhaustivity}

There is an upper bound on $k$ such that the Markov graph is assuredly connected and the underlying walk is exhaustive/irreducible. In particular, given two graphs $G_1$ and $G_2$ of $\G_\C$, 
there always exists a permutation of size at most $M$ (the number of edges) such that $G_1$ is transformed into $G_2$.

\begin{proof}
The $M$ edges of $G_1$ can be written as 
$\left\lbrace (a_1,b_1); (a_2,b_2) ;...; (a_M,b_M) \right\rbrace $.
Similarly, in $G_2$, because both $M$ and degree sequence are fixed, we can write that $M$ edges originate from the same family $(a_i)_{i\in\{1,...,k\}}$ to another family $(b'_i)_{i\in\{1,...,k\}}$, \hbox{i.e.} these edges can be written as $\left\lbrace (a_1,b'_1); (a_2,b'_2);...; (a_M,b'_M) \right\rbrace$. Because the degree sequence is fixed, families of nodes $b$ and $b'$ contain exactly the same nodes repeated the same number of times.
Thus, $\sigma$ defined as $(b_1,b_2,...,b_M) \overset{\sigma}{\longrightarrow} (b'_1,b'_2,...,b'_M)$
is then a valid $M$-switch permutation which does transform $G_1$ into $G_2$.
\end{proof}

The number of connected components of the Markov graph is thus a monotonously decreasing function of $k$ converging at most for $k=M$. As increasing $k$ guarantees a better coverage of the Markov graph, the relevance of this method lies essentially in improving the confidence in the random mixing achieved by rewiring procedures --- rather than 
addressing convergence speed issues.\footnote{In practice, increasing $k$ comes however at the price of an increasingly slow convergence of the walk, in terms of switch trials, as detailed in the following subsection on complexity.}

\subsection{Data storage format}

One of the first requirements for the data format is to enable quick random selection of edges and subsequent edge switches, i.e. update of the graph. A straightforward option for drawing random links consists in using an array of edges, and picking a random integer lower or equal to the array size. To store the graph, by contrast, we opt for an adjacency list, especially because the operation of constraint checking often requires to access neighbors of a given node (which is possible in $O(\delta)$, where $\delta$ is the node degree).  
Eventually, we thus maintain and update two data structures: an adjacency list and an array. These two data strutures have a comparable size and are respectively most efficient for link selection and graph operations.

\subsection{Complexity}

Carrying out a $k$-switch in $G\in\G_\C$ consists in:
\begin{enumerate}
\item Finding $k$ random edges in $G$ represented as an adjacency list, {in $O(k)$};
\item $k$-switching their extremities into a resulting graph $G'$, {in $O(k)$};
\item Verifying that $G'$ respects the constraint set, \hbox{i.e.} $G'\in\G_\C$, {in $O(f_{\G_\C})$} related to a given design of the constraint check. 
\end{enumerate}
$\C$ should be such that there exists a tractable check on any graph of $\G_\C$.\footnote{Various optimizations of this very step are open to a discussion which depends on the chosen external set of constraints $\C$, but are obviously outside the scope of the present paper. In particular, we assume that $f_{\G_\C}$ is not \hbox{e.g.} exponential in $N$ or $M$.} 
In best cases when it is possible to check incrementally if $G'\in\G_\C$ relatively to the $k$ switched edges only, $f_{\G_\C}$ at best belongs to $O(k)$. The complexity of doing $n$ trials of $k$-switches is thus at least $O(nk)$.\\
Additionally, target observables have to be computed at regular intervals to monitor their asymptotical convergence. 
Those target observables shall also be chosen to be tractable. 
If, moreover, the observation frequency is chosen to be sufficiently low, constraint checking shall dominate the overall running time.

\begin{algorithm}[H]
  \SetAlgoLined 
  \DontPrintSemicolon
  \SetKwInOut{Input}{input}
  \SetKwInOut{Output}{output}
  \Input{Graph $G_0 = (V_{0},E_{0})$, stored as an array of adjacency lists;
   number of switching trials: $n$ ; size of the switches: $k$;}
  \Output{graph $G$ produced by $n$ attempts of switching;}
  $G = (V,E) \leftarrow G_0$ ; \tcp*[c]{initialization}
  
	\For{$j\leftarrow 1$ \KwTo $n$}{  
		
	draw randomly $k$ different arcs : $ \left\lbrace \left(  a_i \ , \ b_i \right)  \right\rbrace _{i \in I} \in E$ ; \;
	draw randomly $ \sigma$ a permutation of the index set $I$ ; \;
	build the set of swapped arcs $ \left\lbrace \left( a_i \ , \ b_{\sigma (i)} \right) \right\rbrace _{i \in I}$ ; \;	
	$E' \leftarrow E \cup \left\lbrace \left( a_i \ , \ b_{\sigma (i)} \right) \right\rbrace \setminus \left\lbrace \left(  a_i \ , \ b_i \right)  \right\rbrace $ ; \; 
	define $G' = (V,E')$ ;  \;
	define $ \forall  i \in I \ , \ \mathcal{W}_i =\left\lbrace b \ : \ \exists \ (a_i \ , \ b) \in E \right\rbrace \setminus \left\lbrace b_i \right\rbrace $ ; \; 
	
	 \lIf{$\ \forall  i  \ , \  a_i \neq b_{\sigma (i)} $ \tcp*[c]{test no self-loops}
	 {\bf and } $ \forall  i  \ , \  b_{\sigma (i)} \notin \mathcal{W}_i  $ \tcp*[c]{test no multiple arcs}
	 {\bf and } $ G'\in \G_{\C^+}$  \tcp*[c]{test constraint $ \C^+$}
	  }{
	$G \leftarrow G'$ ; \;}
    }
  \caption{\label{fig:algo}Pseudocode of the $k$-switching procedure in the case of a directed network with constraints: degree distributions, no self-loops, no multiple arcs and a set of constraints $ \C^+$ (associated to the set $\G_{\C^+}$), which depends on the context.}
\end{algorithm}

The reason why large values of $k$ are not necessarily advisable actually lies in the possibility of $k$-switch \emph{failures}, \hbox{i.e.} such that the resulting graph does not anymore belong to $\G_\C$ and thus the walk stays on the same graph at the next step.
Odds of such failure depend in a complicated way on $k$:
on  one hand, when increasing $k$ we are allowing new types of switches, therefore allowing access to possibly more graphs from a given graph of $\G_\C$.  On the other hand, many of these new possible $k$-switches are also likely to fail (\hbox{i.e.} fall on a graph which does not belong to $\G_\C$), because they alter more deeply the graph (\hbox{i.e.} more deeply than $k'$-switches for $k'<k$).  In the end, the proportion of $k$-switch \emph{successes} generally depends on $k$ in a non-monotonous manner.

\emph{In practice}, given an \emph{a priori} fixed number of trials, we observe that the number of successful alterations tends to decrease sharply for large values of $k$ (as shown below \hbox{e.g.} in Tab.
~\ref{res_gcicorr}). In other words, high-order alterations apparently make the walk stay longer on a given graph, although at the same time successful alterations reshuffle more strongly the graph. Put shortly, with increasing $k$, the walk is more likely to stagnate, but when it does not, it is more likely to lead to more different graphs.

\subsection{Random graph sampling using $k$-switches}
It is therefore hard to assess whether the mixing achieved by a $k$-switch-based walk of given length is more efficient or not for higher values of $k$. 
However, the number of connected components of the Markov graph is monotonously decreasing with $k$: increasingly connected portions of $\G_\C$ are visited with increasing values of $k$. 
Because of that, it is relevant to propose an asymptotical approach on $k$. 

More precisely, a $k$-switch walk is stopped when some measures on $\G_\C$ apparently plateau to some values.
Starting with the traditional case $k=2$, we thus progressively increase $k$ up to a ``{sufficient}'' value, \hbox{i.e.} such that the measurements appear to plateau from some $k_0$; as is classical in asymptotical convergence of simulation-based methods.
As we will see in the following section, it seems empirically that even very small values of $k$ are often satisfactory.

\section{Illustrations on practical cases}\label{sec:practical}

In addition to the earlier toy example $\C_0$ shown on Fig.~\ref{fig:metagraph} on an extremely small graph, we now illustrate this asymptotical approach on four practical cases for various kinds of constraints. 
For the sake of clarity, we gathered  in Appendix~\ref{app:constraints} the descriptions of constraint checking algorithms and their respective complexity. 
Note that, here, we only consider constraints on graphs without multiple edges; the higher-order switching approach may nonetheless be used in the context of multigraphs.

\subsection{Constraint based on oriented and colored triangles}

We first suggest a quite fictitious constraint $\C_1$ such that:
\begin{quote}\small\em
\begin{enumerate}[(i)]
\item $C_1^\emptyset$: the graph is directed and made of $N$ nodes, each one having one outgoing and one incoming arc;
\item $C_1^+$:
\begin{itemize}
\item nodes are equally divided into $3$ groups of $N/3$ nodes, each denoted with a color: red (R), green (G), or blue (B);
\item the graph is made of $N/3$ isolated and oriented cycles of $3$ nodes (\hbox{i.e.} $N$ isolated triangles such that each node points to a single other node of the triangle).
\end{itemize}
\end{enumerate}
\end{quote}

\begin{figure}
\begin{minipage}{0.45\linewidth}
\begin{center}
\includegraphics[width=.9\linewidth]{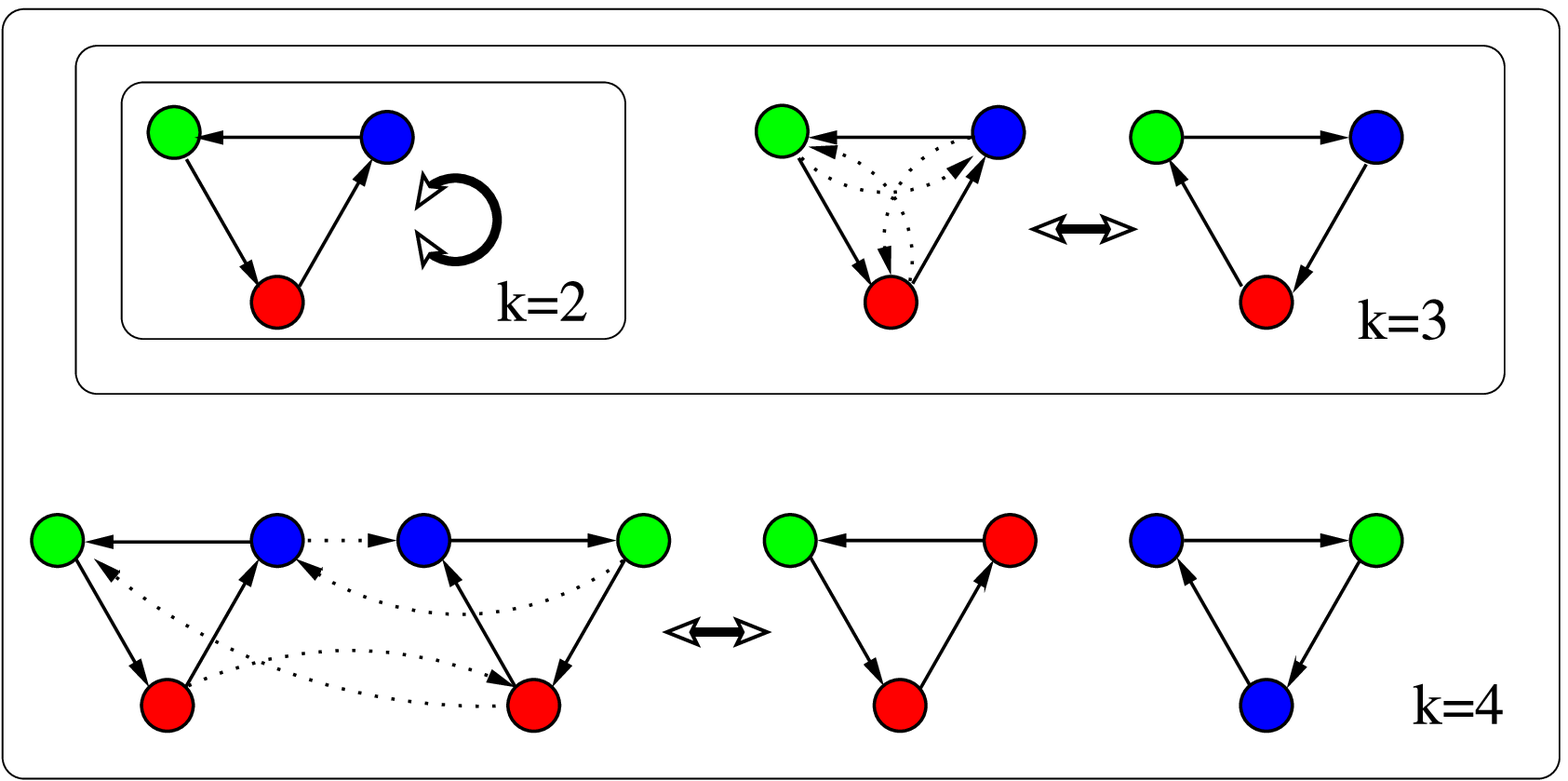}
\end{center}
\end{minipage}
\begin{minipage}{0.5\linewidth}
\begin{center}
\includegraphics[width=.9\linewidth]{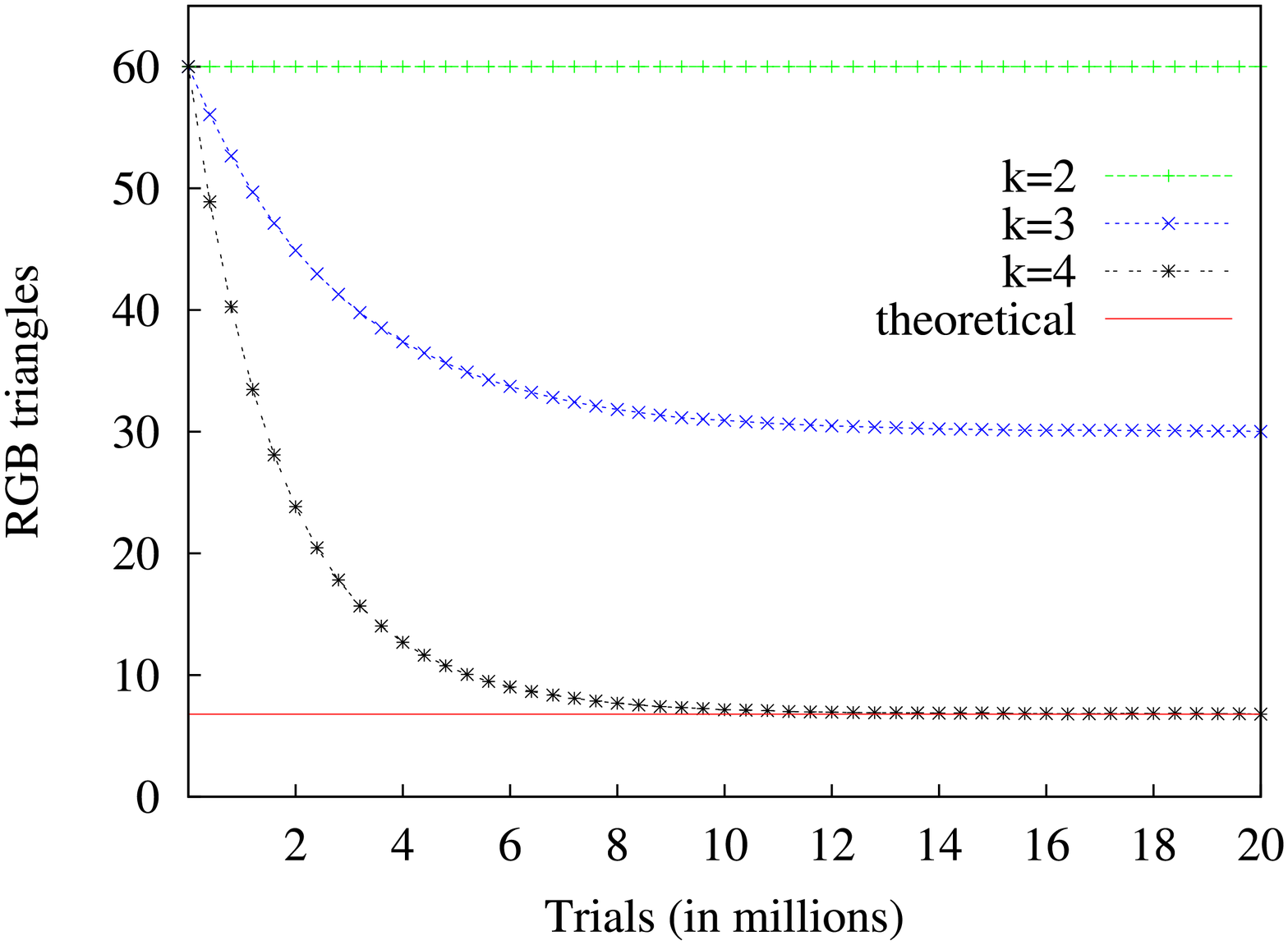}
\end{center}
\end{minipage}
\caption{\emph{Left:} Illustration of the increasing possibilities of $k$-switches for $k\in\{2,3,4\}$ in the case of ``R-B-G'' triangles. \label{fig:swRVBk3}\label{fig:swRVBk4}
\emph{Right:} Number of ``R-B-G'' triangles with respect to the number of $k$-switch trials, for $k\in\{2,3,4\}$ (averages and corresponding confidence intervals computed over 10\,000 simulations for each $k$). \label{fig:cvg}
}
\end{figure}

Graphs of $\G_{\C_1}$ are thus such that each node exactly has an in-degree of 1 and an out-degree of 1. Suppose we want to randomly draw an element of $\G_{\C_1}$ using $k$-switches, starting with an initial graph $G_0$ such that each triangle is ``R-G-B-oriented'', \hbox{i.e.} a red node points to a green one which points to a blue one which points to the red one.

For $k=2$, the only possible $k$-switch is identity, so that in the Markov graph it is not possible to leave $G_0$.  For $k=3$, possible $k$-switches reshuffle links within a given triangle, as illustrated on Fig.~\ref{fig:swRVBk3}; the associated walk can only lead to ``R-G-B-oriented'' and ``R-B-G-oriented'' triangles. For $k=4$, link exchanges are possible between triangles, so that eventually all combinations of colored triangles are possible (including non trichromatic triangles ``R-R-R'', ``R-G-G'', etc.).\footnote{The corresponding Markov graph is thus connected for $k=4$, which hence happens much before $k=M$.
}

\begin{table}[!h]
\begin{center}
\scriptsize
\caption{
Proportion of triangles of each type with respect to $k$, averaged over 10000 completed simulations consisting of $10^8$ trials, including the respective mean number of effectively successful $k$-switches. The last column features the theoretical average value over all graphs of $\G_{\C_1}$. \label{res_rvb}
}
{
\begin{tabular}{c||c|c|c|c|c||c}
& $k=2$ & $k=3$ & $k=4$ & $k=5$ & $k=6$ & \em Theoretical $\langle\G_{\C_1}\rangle$ \\
\hline
\bf R-R-R & 0. & 0. & 0.036 & 0.036 & 0.036 & 0.036\\
\bf  G-G-G & 0. & 0. & 0.036 & 0.036 & 0.036 & 0.036\\
\bf  B-B-B & 0. & 0. & 0.036 & 0.036 &  0.036 & 0.036\\
\bf  R-G-G & 0. & 0. & 0.111 & 0.111 & 0.111 & 0.111\\
\bf  R-B-B & 0. & 0. & 0.111 & 0.111 & 0.111 & 0.111\\
\bf  G-G-B & 0. & 0. & 0.111 & 0.111 &  0.111 & 0.111\\
\bf  G-B-B & 0. & 0. & 0.111 & 0.111 & 0.111 & 0.111\\ 
\bf  R-R-B & 0. & 0. & 0.111 & 0.111 & 0.111  & 0.111\\
\bf R-R-G & 0. & 0. & 0.111 & 0.111 & 0.111 & 0.111\\
\bf  R-B-G & 0. & 0.500 & 0.113 & 0.113 & 0,113 &  0.113\\
\bf R-G-B & 1.000 & 0.500 & 0.113 & 0.113 & 0,113 & 0.113\\
\hline
\em Successes & 0 & $997 \pm 74$ & $2643 \pm 108$ &  $2067 \pm 132$ & $936 \pm 55 $ & - \\
\end{tabular}
}
\end{center}
\end{table}

Considering a trivial target observable which is the proportion of triangles of a given color-orientation, we now compare the performance of $k$-switch-based walks for $k\in\{2,3,4,5,6\}$.  Using simulations on graphs of $N=180$ nodes, we consider the plateauing values of each walk, as shown on Fig.~\ref{fig:cvg}. We then gather in Tab.~\ref{res_rvb}  the various averages of such values obtained over 10000 simulations for each $k$.
We see that average values plateau for $k=4$ which generally fits well the theoretical values, which can be analytically computed for $\C_1$ (see also Tab.~\ref{res_rvb}). 
\rev{However, values obtained for $k=2$ (classical \emph{S\&H}) and $k=3$ are significantly different from the theoretical values, indicating that the corresponding Markov processes are unable to reach every graph of the set $\G_{\C_1}$. 
In particular, the classical \emph{S\&H} method cannot be used in the case of $\C_1$ to generate a random sample, whereas the multiple edges switching method with $k \geq 4$ is reliable.}

Such apparently arbitrary constraints can actually be relevant when considering \hbox{e.g.} complex molecular edifices modeled as graphs linking molecules according to chemical constraints \cite{ralaivola2005graph}.

\subsection{Constraint based on correlations of degrees}

\rev{We now consider constraint $\C_2$ imposing that:
\begin{quote}\small\em
\begin{itemize}
\item $\C_2^\emptyset$: the graph is directed, without self-loops nor multiple edges and has a fixed degree sequence,
\item $\C_2^+$: the distribution of out-degree correlations between pairs of connected nodes is fixed. In other words, the number of links connecting nodes of some out-degree to nodes of some (possibly distinct) out-degree remains the same across the set of graphs.
\end{itemize}
\end{quote}}

\begin{figure}[!h]
\begin{center}
\includegraphics[width=0.5\linewidth]{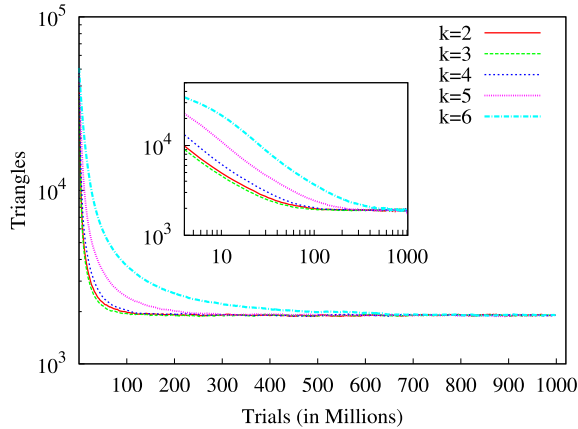}
\end{center}
\caption{Number of directed triangles with respect to the number of $k$-switch trials ($k\in[2,6]$).\label{fig:convergence}}
\end{figure}

\rev{The practical interest of this constraint becomes explicit in the empirical case of a hyperlink  citation network. In qualitative terms, this constraint should in effect help in appraising how much correlations in citing activities (in terms of out-degrees) explain the existence of cyclic citation patterns (in terms of directed triangles). To this end,} we start with an initial graph $G_0$ extracted from the 50,000 first web pages from the network database used in \cite{albe:diam}\footnote{ Available from http://www.barabasilab.com/rs-netdb.php}, we denote this database {\em WWW}. 
We carry out one billion trials in each walk corresponding to $k$-switches for $k\in[2,6]$. We measure the average number of directed triangles (\hbox{i.e.} oriented cycles of length 3) of graphs of $\G_{\C_2}$ thereby estimating how much $\C_2$ contributes to this kind of topological patterns. 
Results are gathered on Tab.~\ref{res_gcicorr} and Fig.~\ref{fig:convergence}.

\newcommand{\expo}[1]{${\cdot10^{#1}}$}

\begin{table}[!h]
\begin{center}
\footnotesize
\caption{
Number of directed triangles 
 with respect to $k$, averaged over 50 completed walks consisting of 1 billion trials, and respective number of effectively successful $k$-switches. Standard deviation are generally negligible and never exceed 5\% of the observed mean. \label{res_gcicorr}
}
{\begin{tabular}{c||c|c|c|c|c|c}
\multirow{2}{1.5cm}{\textit{Target observables}} &\multirow{2}{*}{Starter $G_0$} &\multirow{2}{*}{$k=2$}  &\multirow{2}{*}{$k=3$}  &\multirow{2}{*}{$k=4$}  & \multirow{2}{*}{$k=5$} &\multirow{2}{*}{$k=6$} \\
&&&&&&\\
\hline
\includegraphics[width=0.38cm]{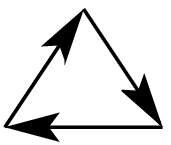}
& 50.77\expo{3}
& 1.92\expo{3}
& 1.91\expo{3} 
& 1.91\expo{3} 
& 1.92\expo{3} 
& 1.91\expo{3} 
\\

\includegraphics[width=0.38cm]{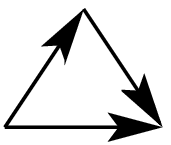}
& 31.70\expo{4}
& 2.90\expo{4} 
& 2.88\expo{4} 
& 2.89\expo{4} 
& 2.90\expo{4} 
& 2.88\expo{4}
\\

\includegraphics[width=0.38cm]{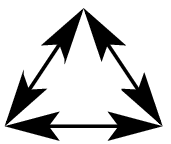}
& 15,423
& 59
& 56
& 58
& 58
& 59 

\\

\hline
\em Successes  
& - 
& 6.96\expo{7}
& 8.22\expo{7}
& 5.28\expo{7}
& 2.50\expo{7}
& 1.00\expo{7} \\
\end{tabular}
}
\end{center}
\end{table}

In spite of their diverse convergence speeds and success rates, $\forall k\in\{2,3,4,5,6\}$ walks converge to a same average number of such directed triangles. As is usually the case with random graphs with constraints, and contrarily to the previous example, we are trying to empirically estimate the theoretical average of this measure on $\G_{\C_2}$. We therefore assume that the plateauing of limit measures for increasing $k$ is a sufficient indication that this empirical estimate can be trusted, which is classical with simulation-based convergence --- similarly, the user may also decide to extend simulations to higher values of $k$.
These results suggest that the reshuffling process is likely to cover well $\G_{\C_2}$ even for $k=2$, \hbox{i.e.} traditional edge swaps. As such, the $k$-switch approach provides an increasing confidence in the simulation estimate of this measure. Qualitatively, because average observable values for $\G_{\C_2}$ do not match those of $G_0$, we have additional confidence in the interpretation that correlations in citation activities does not suffice to explain cyclic citation patterns.\\

To get some insights on how the convergence process varies with  input size,
we implement the algorithm on smaller samples of this dataset: the first 20,000 and 10,000 pages.
Corresponding results are gathered on Table~\ref{tab:res_corr}, providing information about computational requirements in the various cases\footnote{Computations have been made using a standard computer (2x2.33GHz processor, 2GB memory).}.
As will also be the case in the next examples, it seems to be difficult to find any obvious relationship between input size and the number of trials necessary to observe the convergence.

\begin{table}[h!]
\begin{center}
\footnotesize
\caption{
\label{tab:res_corr}Experimental values obtained for constraint $\C_2$ on different inputs (with $N$: number of nodes, $M$: number of arcs): minimum $k$ measured to obtain a uniformly random sample, approximate amount of trials needed for convergence, maximum memory space used during the process.
}
{\begin{tabular}{c|c|c|c|c|c}
\textit{Input} & $N$ & $M$ & $k$ threshold & approximate number of trials & memory used  \\
\hline
{\em WWW-50K} & 50,000 & 143,592 & $2$ & $\sim 1000$m & 13 MB \\
{\em WWW-20K} & 20,000 & 63,224 & $2$ &  $\sim 250$m & 8 MB \\
{\em WWW-10K} & 10,000 & 36,970 & $2$ &  $\sim 250$m & 5 MB \\
\end{tabular}
}
\end{center}
\end{table}

\subsection{Constraint based on triangles}

As said above, it is straightforward to apply the method with constraints on undirected graphs. $\C_3$, and $\C_4$ below, are of this kind.

$\C_3=\C_3^\emptyset\cup\C_3^+$ is such that:
\begin{quote}\small\em
\begin{itemize}
\item $\C_3^\emptyset$: the graph is undirected, with a fixed degree distribution, has no multiple edges nor self-loops
\item $\C_3^+$: the number of (undirected) triangles remains the same.
\end{itemize}
\end{quote}

The interest of $\C_3$ can be illustrated in the case of a collaboration network.
The amount of distinct motifs of size four will be our target observables. In that case, $\C_3$ practically aims at checking whether the size and shape of the close neighborhood of a scientist in this field is related to the cohesiveness between agents --- that is, more precisely, to check how the tendency to do triangular interactions influences the number and connectedness of neighbors at distance 1 and 2.

$G_0$ is an undirected graph of collaborations between scientists extracted from the Anthropological Index Online database.\footnote{Available from http://aio.anthropology.org.uk/aiosearch} 
The dataset we use focuses on a specific subfield consisting of Scandinavian archeology-related papers published over the period 2000--2009: 
nodes are paper authors, links feature collaborations between authors in these papers. $G_0$ contains $273$ individuals and $280$ links.

\begin{figure}[h!]
\begin{center}
\includegraphics[width=0.5\linewidth]{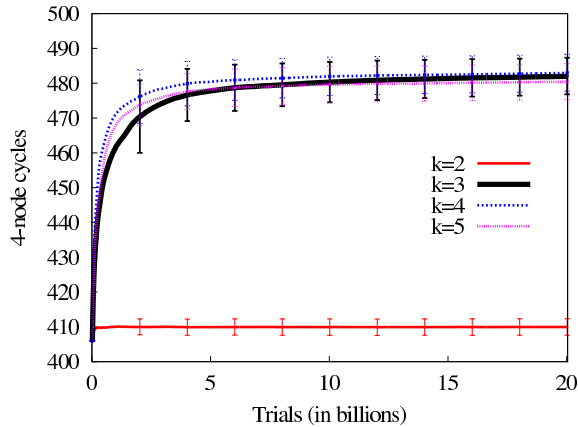}\end{center}
\caption{\label{fig:asympt}Cumulative mean number of 4-nodes cycles for $\C_3$.}
\end{figure}

Results of the corresponding exploration of the random graph space defined by $\C_3$ are gathered on Fig.~\ref{fig:asympt} and Tab.~\ref{fig:aio} for motifs of size four, for which there is significant variation from $G_0$ for $k>2$.  More importantly, these diverging results do not appear when using $k=2$, but only appear from $k>2$, being then similar for all $k\in\{3,4,5,6\}$. 
Thus, the usual \emph{S\&H} method --- unlike the generalized switching method with $k \geq 3$ --- cannot be used to generate a uniformly random subset of $\G_{\C_3}$ on this particular dataset: the obtained sample would be significantly biased.
In other words, only by going beyond $k=2$ makes it possible to show that $\C_3$ is not sufficient to explain the particular shape of the neighborhood of these agents in this empirical network.

\begin{table}[h!]
\begin{center}\footnotesize
\caption{\label{fig:aio}Mean number of motifs of size four after 20 simulations of 10 billion trials on $G_0$ from the AIO database.}
{\begin{tabular}{c||c|c|c|c|c|c}
\textit{Target observables} & Starter $G_0$ & $k=2$ & $k=3$ & $k=4$ & $k=5$ & $k=6 $ \\
\hline
\includegraphics[width=0.4cm]{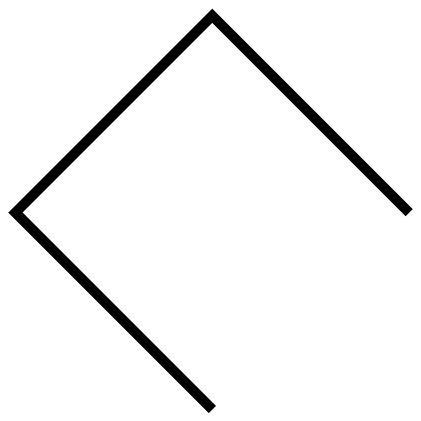} 
& 2794 &  2799 $\pm$ 4 & 2907 $\pm$ 53 & 2933 $\pm$ 32 & 2942 $\pm$ 64 & 2894 $\pm$ 42\\
\includegraphics[width=0.4cm]{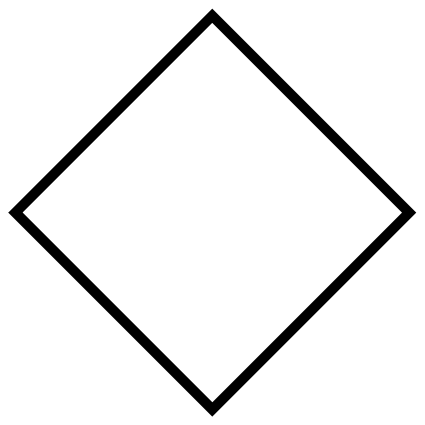}
& 406 & 410 $ \pm $ 3 & 483 $ \pm $ 6 & 483 $ \pm $ 5 & 481 $ \pm $ 5 & 482 $ \pm $ 6\\
\includegraphics[width=0.4cm]{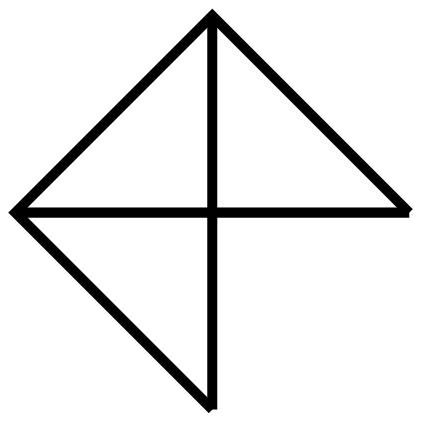}
& 730 & 734 $ \pm $ 3 & 843 $ \pm $ 9 & 841 $ \pm $ 10 & 841 $ \pm $ 6  &  840 $ \pm $ 8\\
\includegraphics[width=0.4cm]{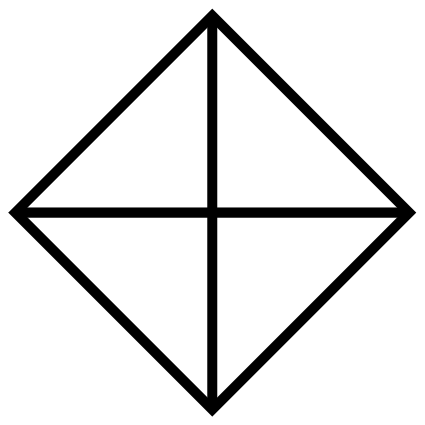}
& 108 & 108 $ \pm$ 0  & 120 $ \pm$ 2 & 120 $ \pm$ 2 & 120 $ \pm$ 2 & 119 $ \pm$ 2\\
\hline
\em Successes (in millions) &  - & 79 & 166 & 96  & 34  & 8\\
\end{tabular}
}
\end{center}
\end{table}

On Table~\ref{tab:res_tri} we gather results on the convergence process on larger collaboration databases extracted from the AIO database in other geographical area, namely the British Isles and the whole of Europe, over the same period of time. Qualitative results on the relationship between $\C_3$ and target observables hold, yet there is, again, no obvious relationship between convergence and input size \& type.  

\begin{table}[h!]
\begin{center}
\footnotesize
\caption{
\label{tab:res_tri}Experimental values obtained for constraint $\C_3$ on different inputs.
}
{\begin{tabular}{c|c|c|c|c|c}
\textit{Input} & $N$ & $M$ & $k$ threshold & approximate number of trials & memory used  \\
\hline
{\em Scandinavia} & 273 & 280 & $3$ & $\sim 20,000$m & 2 MB \\
{\em British Isles} & 807 & 1020 & $2$ & $\sim 10,000$m &  2 MB \\
{\em Europe} & 12112 & 9090 & $2$ & $\sim 100,000$m & 3 MB \\
\end{tabular}}
\end{center}
\end{table}

\subsection{Constraint based on connected components}

Finally, $\C_4$ addresses the issue of connected components. $\C_4$ is such that:
\rev{
\begin{quote}\small\em
\begin{itemize}
\item $\C_4^\emptyset$: the graph is undirected, with a fixed degree distribution, has no multiple edges nor self-loops
\item $\C_4^+$: distribution of the sizes of connected components remains the same
\end{itemize}
\end{quote}}

$G_0$ is an undirected graph built from human metabolic pathways listed in the Biocyc database\footnote{http://www.biocyc.org}: each node is a protein, and each link connects any two proteins involved in the same biochemical pathway.
It features 679 nodes and  11\,030 links. $\C_4$ aims at checking whether the existence of islands of pathways, as represented by connected components, is correlated with the presence of particular local, short-range interactions patterns between specific proteins.

Simulation results are featured on Tab.~\ref{fig:bpw}: averages of statistical variables obtained over corresponding explorations of $\G_{\C_4}$ do not match those of $G_0$. This suggests that $\C_4$ is not a possible explanation for the presence of 3- and 4-sized local patterns in this metabolic pathway network.  

In this case, going beyond $k=2$ did not yield any particular improvement on the random mixing process results, yet provided a stronger confidence on the random exploration of $\G_{\C_4}$.

\begin{figure}[h!]
\begin{center}
\includegraphics[width=.5\linewidth]{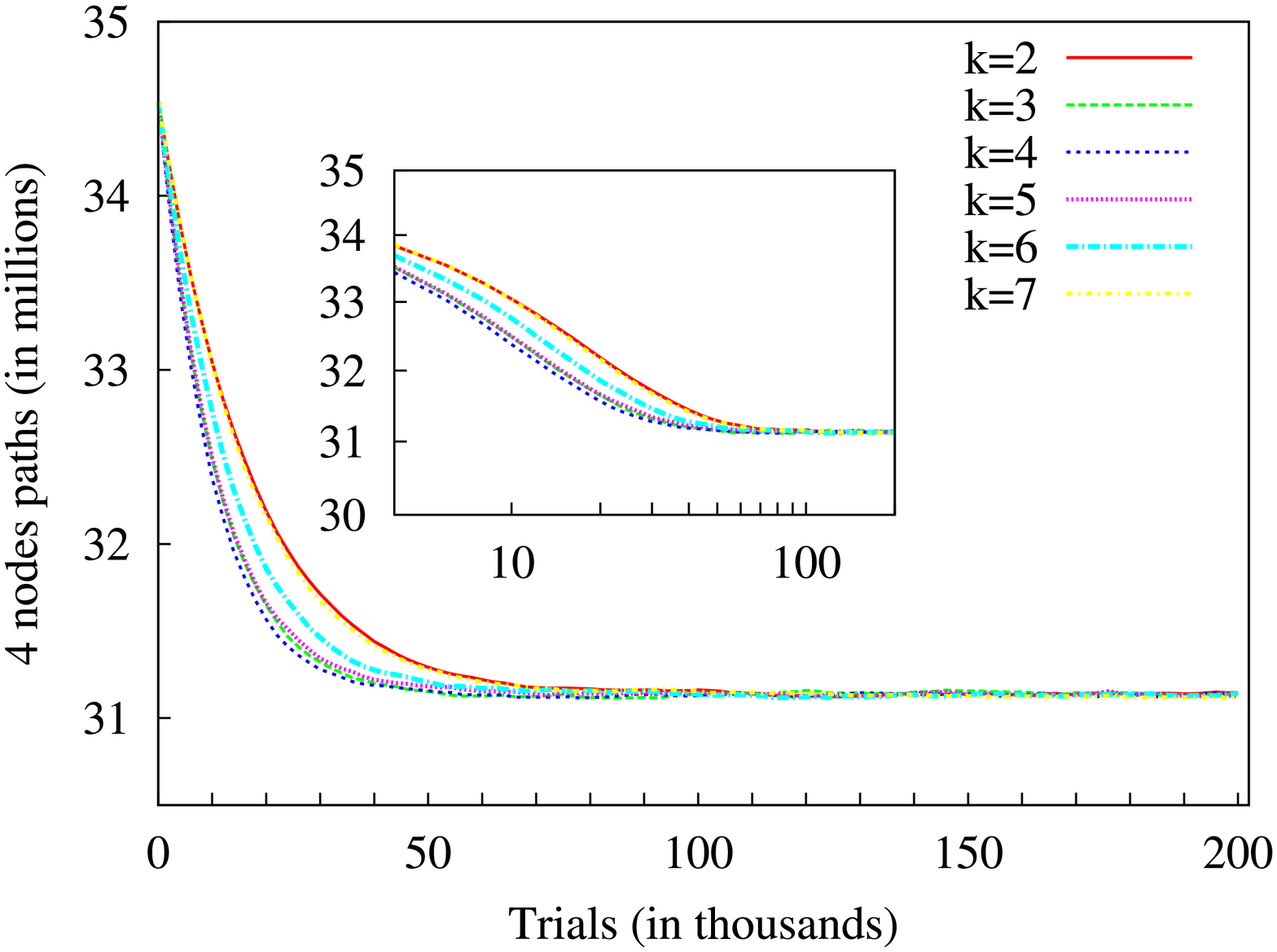}\end{center}
\caption{\label{fig:compo} 
Number of undirected 4-nodes paths with respect to the number of $k$-switch trials ($k \in \left[ 2,7\right] $) for $\C_4$.}
\end{figure}

\begin{table}[h!]
\begin{center}\footnotesize
\caption{\label{fig:bpw}Mean number of patterns of size 3 and 4 on 50 simulations involving $200\,000$ trials on $G_0$ for 'Pathways'.}
{\begin{tabular}{c||c|c|c|c|c|c|c}
\textit{Target observables} & Starter $G_0$ & $k=2$ & $k=3$ & $k=4$ & $k=5$ & $k=6 $ & $k=7$ \\
\hline
\includegraphics[width=0.4cm]{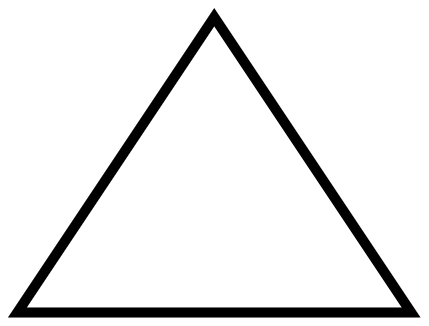}
& 161.3 \expo{3}
& 51.7 \expo{3}
& 51.7 \expo{3} 
& 51.7 \expo{3}
& 51.7 \expo{3}
& 51.7 \expo{3}
& 51.7 \expo{3}\\
\includegraphics[width=0.4cm]{4_clique.eps}
& 2070 \expo{3} 
& 178 \expo{3}
& 178 \expo{3}
& 178 \expo{3}
& 178 \expo{3}
& 178 \expo{3} 
& 177 \expo{3}\\
\includegraphics[width=0.4cm]{4_path.eps}
& 34.5 \expo{6} 
& 31.1 \expo{6}
& 31.1 \expo{6}
& 31.1 \expo{6}
& 31.1 \expo{6}
& 31.1 \expo{6}
& 31.1 \expo{6}\\
\hline
\em Successes  & - & 42,300 & 60,400 & 52,800 & 38,100  & 25,500 & 20,400\\
\end{tabular}}
\end{center}
\end{table}

Again, we run the algorithm on other network datasets: biochemical pathways of {\em Aquifex aeolicus} (denoted {\em aaeo}) and {\em Burkholderia pseudomallei} ({\em bpse}), see Table~\ref{tab:res_compo}. Qualitative results hold too, while there is still no obvious relationship between convergence features and input size \& type.

\begin{table}[h!]
\begin{center}
\footnotesize
\caption{\label{tab:res_compo}Experimental values obtained for constraint $\C_4$ on different inputs.}
{\begin{tabular}{c|c|c|c|c|c}
\textit{Input} & $N$ & $M$ & $k$ threshold & approximate number of trials & memory used  \\
\hline
{\em aaeo} & 264 & 1,193 & $2$ & $\sim20,000$ & 2 MB \\
{\em Human} & 679 & 11,030 & $2$ &  $\sim200,000$ & 3 MB \\
{\em bpse} & 1,447 & 20,620 & $2$ & $\sim500,000$ & 6 MB \\
\end{tabular}}
\end{center}
\end{table}

\section*{Conclusion}
Pairwise edge swapping methods, such as \emph{S\&H}, are relevant to generate uniformly random samples of graphs in some simple cases, such as degree distributions. 
As constraints get 
stronger than just degree distributions, pairwise edge swaps may not be appropriate since the corresponding Markov graph is likely to be disconnected. We therefore proposed a higher-order switching method, denoted ``$k$-edge switching'', which makes it possible to tackle this issue by improving progressively the connectedness of the Markov graph of the corresponding walk.

While this approach guarantees that it is theoretically possible to navigate uniformly throughout the whole Markov graph for some value of $k$, for high values of $k$ the process is likely to be empirically less and less practicable. As such, this approach nonetheless constitutes an easily implementable method to incrementally explore larger portions of the Markov graph; thereby obtaining an increasing, asymptotically certain confidence on the representativeness of the obtained sample.  
{In particular, this method potentially generates random graphs for any type of constraint preserving degree distributions.  It also makes it possible to incrementally check the robustness of results obtained using traditional edge swaps with $k=2$ (which have no reason to yield valid results as such), thereby improving the confidence on the Markov graph exploration achieved by $2$-switches.}

Put simply, when average measurements on the reshuffled graphs tend to plateau for some successive values of $k$, we suggest that it is empirically sensible to assume that the walk covers a reasonably representative portion of the graph set $\G_\C$ 
--- as such constituting a useful extension of edge swapping random graph generation approaches. 
In this respect, an interesting perspective for the present work would be to find classes of constraints $\C$ for which some low values of $k$ guarantee the connectedness of the $k$-switch Markov graph.

\paragraph{Acknowledgements} {\small We are grateful to Cl\'emence Magnien and Fernando Peruani for interesting discussions, thank the anonymous reviewers for their constructive feedback, and acknowledge useful comments from Hugues Chat\'e and Niloy Ganguly.
This work was partly supported by the Future and Emerging Technologies programme FP7-COSI-ICT of the European Commission through project QLectives (grant no.: \texttt{231200}) and by the French ANR through grant ``Webfluence'' \#ANR-08-SYSC-009.} 

\section*{APPENDIX: Constraint checking algorithms and complexities}
\label{app:constraints}

In this Appendix, we describe briefly some possible algorithms for implementing tests corresponding to the above-described constraints.

\subsection*{Constraints $\C_1$ and $\C_3$}

Constraint $\C_1$ may be implemented by testing whether a switch trial creates as many triangles as it destroys.  For each arc $(a_i,b_i)$ involved in a switch trial, we may list which oriented triangles are being created and destroyed by looking for the out-neighbors of $b_i$ which are also in-neighbors of $a_i$ before and after the switch trial. The same goes for $\C_3$, except for the fact that triangles are not oriented.

A random link has a probability proportional to $\delta$ to be connected to a node of degree $\delta$, and we have to go through the list of neighbors for each neighbor of $b_i$.  The same goes with $a_i$, so that the comparison of both lists of neighbors has eventually an average complexity in $O(\overline{\delta}^4)$, where $\overline{\delta}$ is the mean degree. This yields an overall complexity in $O(nk \overline{\delta}^4)$, where $n$ is the number of trials. Note that $\overline{\delta}$ is always equal to 1 in the case of $\C_1$.

\subsection*{Constraint $\C_2$}

The test corresponding to this specific constraint can be implemented as follows: 
after storing at the beginning of the process the out-degree of each node,
the user checks at each trial that for any couple of degrees ($\delta_1 , \delta_2$), 
links whose extremities have degrees $\delta_1$ and $\delta_2$ are created and destroyed in equal numbers.  This specific test can be done in constant time, yielding an overall time complexity of the algorithm in $O(nk)$.

\subsection*{Constraint $\C_4$}

A very simple (yet not optimal) way to implement this constraint test is to check, for each link involved in a switch, the size of the connected component it belongs to before and after the switch.
This can be done in $O(M)$ by using a breadth first search algorithm. This induces a global complexity in $O(nkM)$.

\bibliographystyle{acmsmall}
\bibliography{biblio}

\begin{thebibliography}{}

\bibitem[\protect\citeauthoryear{Albert, Jeong, and Barabasi}{Albert
  et~al\mbox{.}}{1999}]{albe:diam}
{\sc Albert, R.}, {\sc Jeong, H.}, {\sc and} {\sc Barabasi, A.-L.} 1999.
\newblock Diameter of the world wide web.
\newblock {\em Nature\/}~{\em 401}, 130--131.

\bibitem[\protect\citeauthoryear{{Artzy-Randrup} and {Stone}}{{Artzy-Randrup}
  and {Stone}}{2005}]{PhysRevE.72.056708}
{\sc {Artzy-Randrup}, Y.} {\sc and} {\sc {Stone}, L.} 2005.
\newblock Generating uniformly distributed random networks.
\newblock {\em PRE\/}~{\em 72,\/}~5, 056708.

\bibitem[\protect\citeauthoryear{Bansal, Khandelwal, and Meyers}{Bansal
  et~al\mbox{.}}{2008}]{bansal2008ecr}
{\sc Bansal, S.}, {\sc Khandelwal, S.}, {\sc and} {\sc Meyers, L.} 2008.
\newblock {Evolving Clustered Random Networks}.
\newblock {\em Arxiv preprint cs.DM/0808.0509\/}.

\bibitem[\protect\citeauthoryear{Bender and Canfield}{Bender and
  Canfield}{1978}]{bender1978anl}
{\sc Bender, E.} {\sc and} {\sc Canfield, E.} 1978.
\newblock {The asymptotic number of labeled graphs with given degree
  sequences}.
\newblock {\em J. Combin. Theory Ser. A\/}~{\em 24,\/}~3, 296--307.

\bibitem[\protect\citeauthoryear{Colbourn}{Colbourn}{1977}]{colbourn1977gg}
{\sc Colbourn, C.} 1977.
\newblock {\em {Graph generation}}.
\newblock University of Waterloo.

\bibitem[\protect\citeauthoryear{Coolen, De~Martino, and Annibale}{Coolen
  et~al\mbox{.}}{2009}]{coolen2009constrained}
{\sc Coolen, A.}, {\sc De~Martino, A.}, {\sc and} {\sc Annibale, A.} 2009.
\newblock {Constrained Markovian dynamics of random graphs}.
\newblock {\em Journal of Statistical Physics\/}~{\em 136,\/}~6, 1035--1067.

\bibitem[\protect\citeauthoryear{Cooper, Dyer, and Greenhill}{Cooper
  et~al\mbox{.}}{2006}]{cooper2006srg}
{\sc Cooper, C.}, {\sc Dyer, M.}, {\sc and} {\sc Greenhill, C.} 2006.
\newblock {Sampling regular graphs and a peer-to-peer network}.
\newblock {\em Combinatorics, Probability and Computing\/}~{\em 16,\/}~04,
  557--593.

\bibitem[\protect\citeauthoryear{Eggleton}{Eggleton}{1973}]{eggleton1973graphi%
c}
{\sc Eggleton, R.} 1973.
\newblock {Graphic sequences and graphic polynomials: a report}.
\newblock {\em Infinite and Finite Sets\/}~{\em 1}, 385--392.

\bibitem[\protect\citeauthoryear{Feder, Guetz, Mihail, and Saberi}{Feder
  et~al\mbox{.}}{2006}]{feder2006lsm}
{\sc Feder, T.}, {\sc Guetz, A.}, {\sc Mihail, M.}, {\sc and} {\sc Saberi, A.}
  2006.
\newblock {A local switch Markov chain on given degree graphs with application
  in connectivity of peer-to-peer networks}.
\newblock In {\em Proc. of FOCS}. Vol.~6. 69--76.

\bibitem[\protect\citeauthoryear{Gkantsidis, Mihail, and Zegura}{Gkantsidis
  et~al\mbox{.}}{2003}]{gkan:mark}
{\sc Gkantsidis, C.}, {\sc Mihail, M.}, {\sc and} {\sc Zegura, E.} 2003.
\newblock The markov chain simulation method for generating connected power law
  random graphs.
\newblock In {\em Proc. 5th Workshop on Algorithm Engineering and Experiments
  (ALENEX)}.

\bibitem[\protect\citeauthoryear{Guillaume, Latapy, and Le-Blond}{Guillaume
  et~al\mbox{.}}{2005}]{guillaume2005statistical}
{\sc Guillaume, J.}, {\sc Latapy, M.}, {\sc and} {\sc Le-Blond, S.} 2005.
\newblock {Statistical analysis of a P2P query graph based on degrees and their
  time-evolution}.
\newblock {\em Distributed Computing-IWDC 2004\/}, 439--465.

\bibitem[\protect\citeauthoryear{Guruswami}{Guruswami}{2000}]{guruswami2000rmm}
{\sc Guruswami, V.} 2000.
\newblock {Rapidly mixing markov chains: A comparison of techniques}.
\newblock MIT Laboratory for Computer Science.
\newblock Available on cs.washington.edu/homes/venkat/pubs/papers.html.

\bibitem[\protect\citeauthoryear{Kannan, Tetali, and Vempala}{Kannan
  et~al\mbox{.}}{1997}]{kannan1997smc}
{\sc Kannan, R.}, {\sc Tetali, P.}, {\sc and} {\sc Vempala, S.} 1997.
\newblock {Simple Markov-chain algorithms for generating bipartite graphs and
  tournaments}.
\newblock In {\em Proceedings of the eighth annual ACM-SIAM symposium on
  Discrete algorithms}. Society for Industrial and Applied Mathematics
  Philadelphia, PA, USA, 193--200.

\bibitem[\protect\citeauthoryear{Mahadevan, Krioukov, Fall, and
  Vahdat}{Mahadevan et~al\mbox{.}}{2006}]{maha:syst}
{\sc Mahadevan, P.}, {\sc Krioukov, D.}, {\sc Fall, K.}, {\sc and} {\sc Vahdat,
  A.} 2006.
\newblock Systematic topology analysis and generation using degree
  correlations.
\newblock In {\em Proc. SIGCOMM'06}. ACM.

\bibitem[\protect\citeauthoryear{Mikl{\'o}s and Podani}{Mikl{\'o}s and
  Podani}{2004}]{miklos2004rpa-annex}
{\sc Mikl{\'o}s, I.} {\sc and} {\sc Podani, J.} 2004.
\newblock {Randomization of presence-absence matrices: comments and new
  algorithms}.
\newblock {\em Ecology Archives\/}~{\em 85,\/}~1, 86--92.
\newblock Appendix A available on
  http://esapubs.org/archive/ecol/E085/001/appendix-A.htm.

\bibitem[\protect\citeauthoryear{Milo, Kashtan, Itzkovitz, Newman, and
  Alon}{Milo et~al\mbox{.}}{2003}]{milo2003ugr}
{\sc Milo, R.}, {\sc Kashtan, N.}, {\sc Itzkovitz, S.}, {\sc Newman, M.}, {\sc
  and} {\sc Alon, U.} 2003.
\newblock {On the uniform generation of random graphs with prescribed degree
  sequences}.
\newblock {\em Arxiv preprint cond-mat/0312.028\/}.

\bibitem[\protect\citeauthoryear{Newman}{Newman}{2004}]{newman2004coauthorship}
{\sc Newman, M.} 2004.
\newblock {Coauthorship networks and patterns of scientific collaboration}.
\newblock {\em Proceedings of the National Academy of Sciences of the United
  States of America\/}~{\em 101,\/}~Suppl 1, 5200.

\bibitem[\protect\citeauthoryear{Ralaivola, Swamidass, Saigo, and
  Baldi}{Ralaivola et~al\mbox{.}}{2005}]{ralaivola2005graph}
{\sc Ralaivola, L.}, {\sc Swamidass, S.}, {\sc Saigo, H.}, {\sc and} {\sc
  Baldi, P.} 2005.
\newblock Graph kernels for chemical informatics.
\newblock {\em Neural Networks\/}~{\em 18,\/}~8, 1093--1110.

\bibitem[\protect\citeauthoryear{Rao, Jana, and Bandyopadhyay}{Rao
  et~al\mbox{.}}{1996}]{rao1996mcm}
{\sc Rao, A.}, {\sc Jana, R.}, {\sc and} {\sc Bandyopadhyay, S.} 1996.
\newblock {A Markov chain Monte Carlo method for generating random (0,
  1)-matrices with given marginals}.
\newblock {\em Sankhy{\=a}: The Indian Journal of Statistics, Series A\/},
  225--242.

\bibitem[\protect\citeauthoryear{Roberts}{Roberts}{2000}]{roberts2000sms}
{\sc Roberts, J.} 2000.
\newblock {Simple methods for simulating sociomatrices with given marginal
  totals}.
\newblock {\em Social Networks\/}~{\em 22,\/}~3, 273--283.

\bibitem[\protect\citeauthoryear{Sinclair}{Sinclair}{1993}]{sinclair1993arg}
{\sc Sinclair, A.} 1993.
\newblock {\em {Algorithms for random generation and counting: a Markov chain
  approach}}.
\newblock Springer.

\bibitem[\protect\citeauthoryear{Stauffer and Barbosa}{Stauffer and
  Barbosa}{2005}]{stauffer2005ses}
{\sc Stauffer, A.} {\sc and} {\sc Barbosa, V.} 2005.
\newblock {A study of the edge-switching Markov-chain method for the generation
  of random graphs}.
\newblock {\em Arxiv preprint cs.DM/0512.105\/}.

\bibitem[\protect\citeauthoryear{Taylor}{Taylor}{1980}]{tayl:cons}
{\sc Taylor, R.} 1980.
\newblock Constrained switchings in graphs.
\newblock {\em Combinatorial Mathematics\/}~{\em 8}, 314---336.

\bibitem[\protect\citeauthoryear{Taylor}{Taylor}{1982}]{taylor1982scc}
{\sc Taylor, R.} 1982.
\newblock {Switchings constrained to 2-connectivity in simple graphs}.
\newblock {\em SIAM Journal on Algebraic and Discrete Methods\/}~{\em 3}, 114.

\bibitem[\protect\citeauthoryear{Viger and Latapy}{Viger and
  Latapy}{2005}]{viger2005eas}
{\sc Viger, F.} {\sc and} {\sc Latapy, M.} 2005.
\newblock {Efficient and simple generation of random simple connected graphs
  with prescribed degree sequence}.
\newblock {\em Lecture Notes in Computer Science\/}~{\em 3595}, 440.

\end{thebibliography}

\end{document}